\documentclass[aps,twocolumn,showpacs,superscriptaddress,floatfix,letter]{revtex4}
\usepackage{graphicx}
\usepackage{psfrag}
\begin{document}
\newcommand{\beq}{\begin{equation}}
\newcommand{\eeq}{\end{equation}}
\newcommand{\ben}{\begin{eqnarray}}
\newcommand{\een}{\end{eqnarray}}
\newcommand{\bea}{\begin{array}}
\newcommand{\eea}{\end{array}}
\newcommand{\om}{(\omega )}
\newcommand{\bef}{\begin{figure}}
\newcommand{\eef}{\end{figure}}
\newcommand{\leg}[1]{\caption{\protect\rm{\protect\footnotesize{#1}}}}
\newcommand{\ew}[1]{\langle{#1}\rangle}
\newcommand{\be}[1]{\mid\!{#1}\!\mid}
\newcommand{\no}{\nonumber}
\newcommand{\etal}{{\em et~al }}
\newcommand{\geff}{g_{\mbox{\it{\scriptsize{eff}}}}}
\newcommand{\da}[1]{{#1}^\dagger}
\newcommand{\cf}{{\it cf.\/}\ }
\newcommand{\ie}{{\it i.e.\/}\ }

\title{DISTRIBUTION OF RESONANCE WIDTHS AND DYNAMICS OF CONTINUUM COUPLING}

\author{G.~L.~Celardo}
\affiliation{Dipartimento di Matematica e Fisica, Universit\`a Cattolica, via Musei 41, 25121, Brescia, Italy}
\affiliation{I.N.F.N., Sezione di Pavia, Italy}
\author{N. Auerbach}
\affiliation{School of Physics and Astronomy, Tel Aviv University,
Tel Aviv, 69978, Israel}
\affiliation{NSCL and Department of Physics and Astronomy, Michigan State
University, East Lansing, Michigan 48824-1321, USA}
\author{F.~M.~Izrailev}
\affiliation{NSCL and Department of Physics and Astronomy, Michigan State
University, East Lansing, Michigan 48824-1321, USA}
\affiliation{Instituto de F\'{\i}sica, Universidad Aut\'{o}noma de
Puebla, Apartado Postal J-48, Puebla, Pue., 72570, M\'{e}xico}
\author{V.~G.~Zelevinsky}
\affiliation{NSCL and Department of Physics and Astronomy, Michigan State
University, East Lansing, Michigan 48824-1321, USA}

\begin{abstract}

We analyze the statistics of resonance widths in a many-body Fermi system with
open decay channels. Depending on the strength of continuum coupling, such a
system reveals growing deviations from the standard chi-square
(Porter-Thomas) width distribution. The deviations emerge from the process of
increasing interaction of intrinsic states through common decay channels; in the
limit of perfect coupling this process leads to the super-radiance phase
transition. The width distribution depends also on the
intrinsic dynamics (chaotic vs regular). The results presented here are
important for understanding the recent experimental data
concerning the width distribution for neutron resonances in nuclei.

\end{abstract}

\date{\today}
\maketitle

Open and marginally stable quantum systems are of great current interest in
relation to numerous applications in nuclear physics of exotic nuclei, chemical
reactions, condensed matter, astrophysics and quantum informatics. The general
problem can be formulated as that of the signal transmission through a
complicated quantum system. The complexity of theoretical description of such
processes is due to the necessity of a consistent unified theory that would cover
intrinsic structure, especially for many-body systems, along with cross sections of
various reactions.

One of the best and historically advanced examples of the manifestation of the
interplay between intrinsic dynamics and decay channels is given by low-energy
neutron resonances in complex nuclei \cite{BM}. The series of these well
pronounced separated resonances were studied long ago
\cite{porter58,porter60,porterbook} and later gave rise to the ``Nuclear Data
Ensemble" \cite{haq82,sissom82}. Interpreting these resonances as
quasi-stationary levels of the compound nucleus formed after the neutron
capture, agreement was found with predictions of the Gaussian Orthogonal
Ensemble (GOE) of random matrices. With exceedingly complicated wave functions
of compound states, the statistical distribution of their components is close
to Gaussian. The neutron decay implements the analysis of a specific component
related to the channel ``neutron in continuum plus a target nucleus in its
ground state". The neutron width is proportional to the squared amplitude of this
component, and the width distribution then is $\chi^{2}_{\nu}$ with $\nu=1$ as
appropriate for one channel [Porter-Thomas distribution (PTD)].

The recent experiments with improved accuracy \cite{koehler10} give evidence of
significant deviations from the PTD so that the attempts to still use the
$\chi^{2}_{\nu}$ distribution for the fit invariably require $\nu<1$.
A non-pure set of resonances, for example, a sequence of mainly $s$-resonances
contaminated by $p$-wave states, would shift the distribution to
a higher number of degrees of freedom, $\nu>1$. The new result
was interpreted as a consequence of an unknown non-statistical mechanism or
just a breakdown of nuclear theory as was claimed in the related article in
``Nature" with a title ``Nuclear theory nudged" \cite{nature10}.

The goal of this letter is to point out that a correct description of unstable
quantum states in a complicated many-body system naturally leads to 
deviations from the GOE and PTD, of the same type as observed in \cite{koehler10}.
The random matrix theory was formulated for local statistics in a closed quantum
system with a discrete spectrum governed by a very complicated Hermitian
Hamiltonian. As such, its predictions were repeatedly checked, qualitatively
and quantitatively, in systems like quantum billiards \cite{nakamura04} and their
experimental embodiment in microwave cavities \cite{stockmann99,alt95}, and in
shell-model calculations for complex atoms \cite{grib94} and nuclei \cite{big}.

However, the presence of open decay channels and therefore the finite lifetime
of intrinsic states unavoidably lead to new phenomena outside of the GOE
framework \cite{verbaarschot85,SZNPA89} as it was clearly demonstrated by the
first numerical simulations \cite{kleinwachter85,mizutori93,izr94}. Two
interrelated effects follow from the fact that we deal with unstable
rather than with strictly stationary states: the level repulsion disappears at
the spacings comparable to the level widths and the growing widths undergo the
redistribution with the trend to collectivization and eventually
formation of super-radiant (short-lived) states along with the narrow (trapped)
states \cite{SZNPA89,SZAP92}. The new dynamics modify the GOE predictions as
well as certain features of Ericson fluctuations \cite{ericsonAP63} in the
regime of overlapping resonances \cite{celardo08}.
The occurrence of a super-radiant transition has been also
demonstrated outside the random matrix theory framework \cite{kaplan}.

A quantum many-body system coupled to open decay channels can be rigorously
described by an effective non-Hermitian Hamiltonian ${\cal H}$ \cite{MW},
\begin{equation}
{\cal H}= H-\,\frac{i}{2}\,W.                               \label{1}
\end{equation}
Here $H$ is the Hermitian Hamiltonian of the closed system that in general includes
virtual (off-shell) coupling to the continuum, while the anti-Hermitian
(on-shell) part $(-i/2)W$ is constructed in terms of the amplitudes ${\bf A}=\{A^{c}_{i}\}$
coupling intrinsic states $|i\rangle$ to the open channels $c$,
\begin{equation}
W={\bf A}{\bf A}^{T} \quad \Rightarrow \quad W_{ij}=\sum_{c({\rm open})}^{M}
A^{c}_{i}A^{c}_{j}.                                   \label{2}
\end{equation}
We consider a time-reversal invariant system, when the (in general depending on
running energy) amplitudes $A^{c}_{i}$ can be taken real. 
The factorized form of
$W$ is an important property that follows from unitarity of 
the scattering matrix
\cite{durand76}. The complex eigenvalues, ${\cal E}=E-(i/2)\Gamma$, 
of the Hamiltonian
(\ref{1}) coincide with the poles of the scattering matrix and 
determine the positions
and the widths of the resonances in cross sections of various reactions.

The factorized matrix $W$ has $M$ non-zero eigenvalues, this number being equal
to that of open channels. 
The matrix $H$ has dimension $N$ that in the nuclear case
should include a large number of shell-model many-body states 
important for the
dynamics in the energy range under consideration; 
in the region of neutron resonances,
$N\sim 10^{5\div 6}\gg M$. With the trace of $W$ equal to $\eta$, the important
parameter defining the dynamics is the ratio of typical ``bare" widths $\eta/N$
of individual states to the energy spacings $D$. 
At small value of this parameter,
an open channel serves as an analyzer that singles out a 
specific component of the exceedingly complicated intrinsic wave function.
Characteristically, the resonance widths in such a system obey the PTD. 
With widths increasing, the system moves in the direction of 
the regime of overlapping resonances.

Let us consider the single-channel case, $M=1$, having in mind the $s$-wave
elastic neutron resonances. With a high level density of intrinsic states at
relevant energy, their local spectral statistics is close to the predictions of
the GOE. Then, essentially owing to the central limit theorem, the individual
components of a typical intrinsic state are Gaussian distributed uncorrelated
quantities, and the neutron widths, being proportional to the absolute
magnitudes of those components, display the PTD. However, the correct
description of the dynamics with the continuum coupling shows the limited
character of this prediction. The imaginary part (\ref{2}) works similar to the
collective multipole forces and creates the interaction between intrinsic
states through continuum. When the coupling is weak,
$\kappa= \eta/ND\ll 1$, we indeed
expect to see well isolated resonances with the PTD of the widths. With growing
continuum coupling (increase of energy from the threshold), the deviations
become more and more pronounced. At $\kappa\sim 1$, a kind of a phase
transition occurs with the sharp redistribution of widths and the segregation
of a super-radiant state accumulating the lion's share of the whole summed
width, an analog of a giant resonance along the imaginary energy axis. The
formal mechanism is clear from the factorized structure of $W$ that, for $M=1$,
has only one non-zero eigenvalue equal to the trace of $W$. Being similar to
super-radiance in optics \cite{dicke54}, 
this mechanism works essentially independently
of the regular or chaotic nature of intrinsic dynamics.

In the case of the GOE-type dynamics of
the closed system and $M=1$, the distribution of complex eigenvalues
can be found analytically. The Ginibre ensemble \cite{ginibre65} of
complex Gaussian matrices is not applicable since it considers
the imaginary parts of the eigenvalues spread over
the complex plane while physical widths are positive.
The exact result, see derivation in \cite{SZNPA89}, is given by
the Ullah distribution \cite{ullah69},
\begin{equation}
{\cal P}(\{E_{n};\Gamma_{n}\})=C_{N}
\prod_{m<n}\,\frac{|{\cal E}_{m}-{\cal E}_{n}|^{2}}{|{\cal E}_{m}-{\cal E}_{n}^{\ast}|}
\prod_{n}\,\frac{1}{\sqrt{\Gamma_{n}}}\,\exp(-NF). \label{3}
\end{equation}
Here $C_{N}$ is a normalization constant; 
the pre-exponential factor describes the
correlations which are reduced to the usual GOE level
repulsion for stable states but for complex energies ${\cal E}_{n}$
contain the interactions with their ``electrostatic images" 
${\cal E}_{n}^{\ast}$.
Along with the Porter-Thomas factor $\Gamma^{-1/2}$, this guarantees that
the widths are positive. 
The ``equilibrium" distribution of complex energies is determined
by their ``free energy",
\begin{equation}
F(\{E_{n},\Gamma_{n}\})= \sum_{n}\left(\frac{E_{n}^{2}}{a^{2}}\,+
\,\frac{\Gamma_{n}}{2\eta}\right)\,+\sum_{m<n}\frac{\Gamma_{m}\Gamma_{n}}{2a^{2}},
                                                         \label{4}
\end{equation}
that includes the interaction between the widths, the last term 
in Eq. (\ref{4}).
The mean level spacing in the
closed system is defined as $D=2a/N$, where $2a$ is the spectral interval of
real energy. The Gaussian ensemble of the decay amplitudes is defined by
the mean values 
$\overline{A_{n}}=0, \;\overline{A_{n}A_{m}}=(\eta/N)\delta_{mn}$. Here,
a regular evolution of the widths as a function of energy of the resonance
is excluded as it is usually done with the rescaling to the reduced widths;
we do not discuss here the way of practical rescaling that may depend on
the specific nucleus. 
Our main purpose here is to show that systematic deviations
from the PTD occur even for the set of reduced widths, and the effects are
caused by the interaction (\ref{4}).

For very small widths, $\kappa \ll 1$, the width interaction is negligible,
the first product in (\ref{3}) reduces to the standard level repulsion, and
the distribution (\ref{3}) is factorized into the product 
of the GOE distribution of real energies and the PTD for the widths. 
While the usual Hermitian perturbation
mixes the intrinsic wave functions and therefore makes their widths close
to each other (level repulsion and width attraction), the anti-Hermitian
interaction through the continuum relaxes the level repulsion but leads to
the collectivization through the common decay channel 
and width repulsion \cite{brentano96}.
In the case of $\kappa \simeq 1$, the width repulsion becomes critical,
and the most probable configuration is the one where this repulsion is
small because the total width is going to concentrate in a single
super-radiant state \cite{SZNPA89,SZAP92}. With the further increase of
$\kappa\gg 1$ (but for a fixed number of open channels), the broad state
becomes a smooth envelope, and we return to the set of $(N-1)$ narrow
resonances.  Below we show the typical evolution of
the width distribution studied in a large-scale numerical simulation.

In the first type of simulations, we considered $H$ as a member of the GOE
\cite{Zel1,VWZ} where the matrix elements are Gaussian random variables,
$\langle H_{ij}^2\rangle =(1+\delta_{ij})/N$. This corresponds to
the limiting case of fully chaotic intrinsic dynamics. In parallel,
we also modeled $H$ by the two-body random ensemble (TBRE) for $n$ fermions
distributed over $m$ single-particle states; the total number of many-body
states is $N=m!/[n!(m-n)!]$; in our simulations $n=6,\;m=12,\;N=924$.
The TBRE is modeled by the Hamiltonian $H=H_0+V$, where the mean field
part $H_0$ is defined by single-particle energies, $\epsilon_j$, with a
Poissonian distribution of spacings and the mean level spacing $d_0$.
The interaction $V$ \cite{FI97} is fixed by the variance of the two-body random
matrix elements, $\langle V_{1,2;3,4}^2\rangle=v_0^2$. While
at $v_0=0$ we have a Poissonian spacing distribution $P(s)$ of
many-body states, for $d_0=0$ (infinitely strong interaction,
$v_0/d_{0} \rightarrow \infty$), $P(s)$ is close to the Wigner-Dyson
distribution typical for a chaotic system. The critical
interaction for the onset of strong chaos is given \cite{FI97} by
$v_{cr}/d_0 \approx 2(m-n)/N_s$, where
$N_s=n(m-n)[1+(n-1)(m-n-1)/4]$ is the number of directly coupled
many-body states in any row of the matrix $H_{ij}$. In
our model, $v_{cr}/d_{0} \approx 1/20$.

We computed the complex eigenvalues of the effective Hamiltonian, Eq.(\ref{1}),
for $10^{3}$ random realizations,
selecting the states at the center of the energy band
where their density is almost constant.

\begin{figure}[h!]
\vspace{0cm}
\includegraphics[width=7cm,angle=-90]{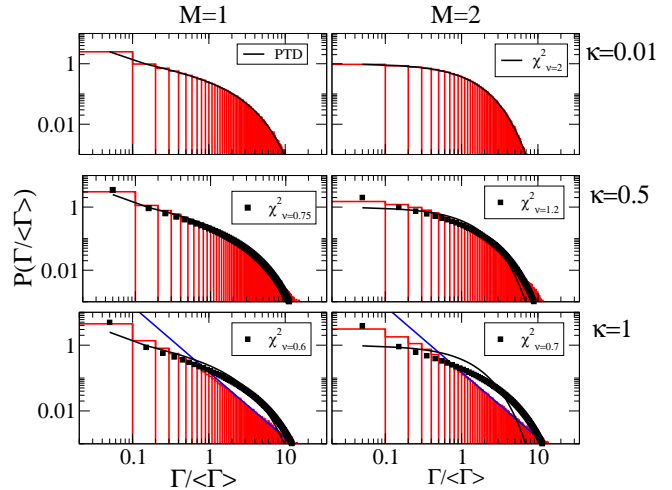}
\caption{ (Color online)
Normalized distribution of the widths for one open channel,
$M=1$, left panels, and for $M=2$, right panels, for the GOE 
intrinsic Hamiltonian and different continuum coupling strength, $\kappa$. 
The numerical results are shown by histograms; 
the expected PTD for $M=1$ and $\chi^2_{\nu=2}$
distribution for $M=2$ are shown by a smooth curve in all panels.
With full squares we show the best fit with a $\chi^2_{\nu}$ distribution.
In the lowest panels, a straight line shows the power law
$P(\Gamma/\langle\Gamma\rangle)\propto(\Gamma/\langle\Gamma\rangle)^{-2}$.
Included are the widths for the resonances in the energy interval $\pm 0.5$ for
$10^3$ different ensemble realizations. }
\label{ptm1}
\end{figure}

The distribution of widths, normalized to their average value,
was obtained for different strengths of continuum coupling.
The numerically obtained distributions were fitted, using
a standard $\chi^2$ test, with a $\chi^2_{\nu}$ distribution for
$\nu$ degrees of freedom, following a common practice in nuclear
data analysis. As a measure of quality of the fit we used the criterion
$\chi^2_r \approx 1$, where $\chi^2_r $ 
is the reduced chi-square value \cite{chi}.
Fig. \ref{ptm1} shows the normalized width distribution for
the GOE case with $M=1$, left panels, and $M=2$, right panels,
for different coupling to the continuum, $\kappa$.
The standard distributions, PTD for $M=1$, and  $\chi^2_{\nu=2}$ for $M=2$,
are valid only when the coupling to the continuum is very weak, while
strong deviations from $\chi^2_{\nu=M}$ appear, both for large and small
$\Gamma$, as we increase the coupling. In Fig. \ref{ptm1} we also show
the best fit possible for a $\chi^2_{\nu}$ distribution; always
the corresponding value of $\nu$ is $<M$, moreover 
the quality of the fit decreases as $\kappa$ increases, see discussion below. 
In the lowest panels in
Fig. \ref{ptm1}, the distribution of the widths is shown at the super-radiance
transition, $\kappa=1$.  The tail here is described by a
power law, see discussion in  Ref. \cite{SFT99},
meaning that no $\chi^2_{\nu}$ distributions would be a good fit.

\begin{figure}[h!]
\vspace{0cm}
\includegraphics[width=7cm,angle=-90]{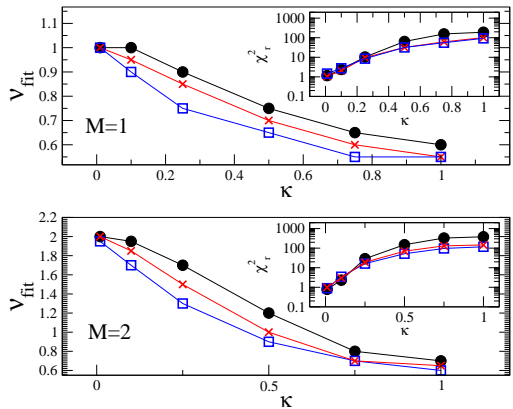}
\caption{ (Color online)
The best fitted value of $\nu$ {\sl vs} the coupling strength to
the continuum, $\kappa$. Full circles refer to the case of GOE
intrinsic Hamiltonian, crosses stand for the case
of TBRE with $v_0=d_0/50$, and squares for the case of
TBRE with $v_0=0$ (no intrinsic chaos); the number of channels is one
(upper panel) and two (lower panel).
The insets show the reduced chi $\chi^2_r$
as a function of $\kappa$. }
\label{nuvsk}
\end{figure}

The dependence of the best fitted value of $\nu$ on the coupling
strength is shown in Fig. \ref{nuvsk} for $M=1$, upper panel,
and for $M=2$, lower panel. Along with
the data for the GOE intrinsic Hamiltonian (circles),
the data for the TBRE intrinsic dynamics are shown by crosses
for the case $v_0=d_0/50$ and by squares for $v_0=0$ (no intrinsic chaos).
Regardless of intrinsic dynamics, the best value of $\nu$
steadily decreases as $\kappa$ increases.
It is clear that a family of $\chi^2_{\nu}$ distributions is not appropriate
to fit our numerical data, except for very weak continuum coupling. Indeed,
the $\chi^2_r$ criterion steeply increases with the coupling strength,
see insets in  Fig. \ref{nuvsk}.
For weak internal chaos, the departure from the  $\chi^2_{\nu=M}$ distribution
is stronger than for chaotic intrinsic dynamics, 
even at weak continuum coupling. 
The absence of intrinsic chaos and corresponding level 
repulsion implies a stronger sensitivity to continuum coupling. 

When analyzing empirical neutron $s$-wave resonances one of the main
difficulties is the $p$-wave contamination. In order to analyze this
problem we considered the case of two non-equivalent channels, with
$10 \%$ of the states coupled to the additional channel by a smaller
coupling strength. When another channel is included,
the best fit value of $\nu$ at weak coupling is always larger than one;
again this value decreases when the continuum coupling is growing.
As expected, we also observed the evolution of the level spacing distribution
$P(s)$ along the real axis, with disappearance of repulsion at spacings
comparable with the level widths \cite{mizutori93}; for example,
at $\kappa=0.5$, $P(0)\approx 0.2$. Interesting physics related to
energy-width correlations will be discussed elsewhere.

To summarize, the normalized width distributions have been analyzed for
a system with $M=1$ or 2 open channels as a function of the continuum coupling
strength. As this coupling increases, the best fit value of $\nu$ for
a  $\chi^2_{\nu}$ distribution decreases below $M$, in accordance to
recent experimental findings \cite{koehler10}. At the same time,
the fit quality becomes poor showing that the standard PTD 
(and in general any $\chi^2_{\nu}$ distribution) is applicable 
only for extremely narrow resonances.
The low-energy neutron resonances in a heavy
nucleus correspond to the very beginning of
the process of width collectivization.
However, already here the deviations from the (GOE$\times$PTD) factorized
distribution are noticeable. These deviations are more pronounced
for regular intrinsic dynamics than for chaotic intrinsic dynamics.
 Therefore, the interpretation
of the width as a strength of the pure neutron component in the 
compound wave function fails due to the coupling through continuum 
that has to be accounted for
in a proper statistical description. The phenomenon under discussion is
of general nature and it may influence all processes of signal transmission
through a quantum system.

We are grateful to V. Sokolov, Y. Fyodorov, and A. Volya for useful comments.
N.A. thanks the NSCL and W. Mittig for hospitality and support;
G.L.C. acknowledges fruitful discussions with F. Borgonovi;
V.Z. acknowledges support from the NSF grant PHY-0758099.
F.I. acknowledge partial support from VIEP BUAP grant EXC08-G.


\begin{thebibliography} {99}

\bibitem{BM} A. Bohr and B.R. Mottelson, {\sl Nuclear
Structure}, Vol. 1 (World Scientific, Singapore, 1998).

\bibitem{porter58} C.E. Porter and R.G. Thomas, Phys. Rev. {\bf
104}, 483 (1956).

\bibitem{porter60} C.E. Porter and N. Rosenzweig, Phys. Rev. {\bf 120}, 1698
(1960).

\bibitem{porterbook} {\sl Statistical Theories of Spectra: Fluctuations},
ed. by C.E. Porter (Academic, New York, 1965).

\bibitem{haq82} R. Haq, A. Pandey, and O. Bohigas, Phys. Rev.
Lett. {\bf 48}, 1986 (1982).

\bibitem{sissom82} D.J. Sissom, J.F. Shriner, Jr., G. Mitchell,
Phys. Rev. Lett. {\bf 48}, 1086 (1982).

\bibitem{koehler10} P.E. Koehler, F. Becvar, M. Krticka, J.A. Harvey,
and K.H. Guber, Phys. Rev. Lett. {\bf 105}, 072502 (2010).

\bibitem{nature10} E.S. Reich, Nature {\bf 466}, 1034 (2010).


\bibitem{nakamura04} K.Nakamura and T.Harayama, {\sl Quantum Chaos and
Quantum Dots: Mesoscopic Physics and Nanotechnology} (Oxford University
Press, 2004).

\bibitem{stockmann99} H.-J. St\"{o}ckmann, {\sl Quantum Chaos:
An Introduction} (Cambridge University Press, 1999).

\bibitem{alt95} H. Alt, H.-D. Gr\"{a}f, H.L. Harney, R. Hofferbert,
H. Lengeler, A. Richter, P. Schardt, and H.A. Weidenm\"{u}ller,
Phys. Rev. Lett. {\bf 74}, 62 (1995).

\bibitem{grib94} V.V. Flambaum, A.A. Gribakina, G.F. Gribakin, and
M.G. Kozlov, Phys. Rev. A {\bf 50}, 267 (1994).

\bibitem{big} V. Zelevinsky, B.A. Brown, N. Frazier, and M. Horoi,
Phys. Rep. {\bf 276}, 85 (1996).

\bibitem{verbaarschot85} J.J.M. Verbaarschot, H.A. Weidenm\"{u}ller,
and M.R. Zirnbauer, Phys. Rep. {\bf 129}, 367 (1985).

\bibitem{SZNPA89} V.V. Sokolov and V.G. Zelevinsky, Nucl. Phys.
{\bf A504}, 562 (1989).

\bibitem{kleinwachter85} P. Kleinw\"{a}chter and I. Rotter, Phys.
Rev. C {\bf 32}, 1742 (1985).

\bibitem{mizutori93} S. Mizutori and V.G. Zelevinsky, Z. Phys.
{\bf A346}, 1 (1993).

\bibitem{izr94} F.M. Izrailev, D. Saher, and V.V. Sokolov, Phys.
Rev. E {\bf 49}, 130 (1994).

\bibitem{SZAP92} V.V. Sokolov and V.G. Zelevinsky, Ann. Phys.
(N.Y.) {\bf 216}, 323 (1992).

\bibitem{ericsonAP63} T. Ericson, Ann. Phys. {\bf 23}, 390 (1963).

\bibitem{celardo08} G.L. Celardo, F.M. Izrailev, V.G. Zelevinsky,
and G.P. Berman, Phys. Lett. B {\bf 659}, 170 (2008);
Phys. Rev. E {\bf 76}, 031119 (2007).

\bibitem{kaplan} G. L. Celardo and L. Kaplan,
Phys. Rev. B {\bf 79}, 155108 (2009); E. Persson, I. Rotter,
H.-J. St\"ockmann, and M. Barth, Phys. Rev. Lett. {\bf 85}, 2478 (2000).

\bibitem{MW} C. Mahaux and H.A. Weidenm\"{u}ller, {\sl Shell Model
Approach to Nuclear Reactions} (North Holland, Amsterdam, 1969).

\bibitem{durand76} L. Durand, Phys. Rev. D {\bf 14}, 3174 (1976).

\bibitem{dicke54} R.H. Dicke, Phys. Rev. {\bf 93}, 99 (1954).

\bibitem{ginibre65} J. Ginibre, J. Math. Phys. {\bf 6}, 3 (1965).

\bibitem{ullah69} N. Ullah, J. Math. Phys. {\bf 10}, 2099 (1969).

\bibitem{brentano96} P. von Brentano, Phys. Rep. {\bf 264}, 57
(1996).

\bibitem{FI97} V.V. Flambaum and F.M. Izrailev, Phys. Rev. E {\bf 56},
5144 (1997).

\bibitem{Zel1} V. V. Sokolov and V. G. Zelevinsky, Phys. Lett.
B {\bf 202}, 10 (1988); Nucl. Phys. {\bf A504}, 562 (1989).

\bibitem{VWZ} J.J.M. Verbaarschot, H.A. Weidenm\"{u}ller, and M.R.
Zirnbauer, Phys. Rep. {\bf 129}, 367 (1985).

\bibitem{chi} Given a histogram with $n$ bins, the reduced
chi squared is defined as $\chi^2_r= 1/(n-1) \sum_{k=_1}^n (O_k-E_k)^2/E_k$,
where $O_k$ is the observed number of events in the $k^{{\rm th}}$ bin and
$E_k$ is the expected number of events.

\bibitem{SFT99} Y.V. Fyodorov and H.-J. Sommers, J.Math. Phys.
{\bf 38}, 1918 (1997); H.-J. Sommers, Y.V. Fyodorov, and M. Titov,
J. Phys. A: Math. Gen. {\bf 32}, L77 (1999).


\end{thebibliography}
\end{document}